\documentclass[12pt,preprint]{aastex}

\def\ltsima{$\; \buildrel < \over \sim \;$}
\def\gtsima{$\; \buildrel > \over \sim \;$}
\def\lsim{\lower.5ex\hbox{\ltsima}}
\def\gsim{\lower.5ex\hbox{\gtsima}}
\def\lapp{\ifmmode\stackrel{<}{_{\sim}}\else$\stackrel{<}{_{\sim}}$\fi}
\def\gapp{\ifmmode\stackrel{>}{_{\sim}}\else$\stackrel{<}{_{\sim}}$\fi}
\def\mcom{M_{\rm COM}}
\def\mpsr{M_{\rm PSR}}

\def\dpsr{d_{\rm PSR}}
\def\rcom{R_{\rm COM}}
\def\rrl{R_{\rm RL}}
\def\Msun{M_{\odot}}
\def\psr{PSR\, J0610$-$2100}
\def\com{COM\, J0610$-$2100}
\newcommand{\vlt}{{\em VLT}}

\newdimen\minuswidth    

\setbox0=\hbox{$-$}

\minuswidth=\wd0

\catcode`@=\active

\def@{\kern\minuswidth}

\setbox0=\hbox{\rm0}

\shorttitle{Optical companion to PSR J0610-2100}

\shortauthors{Pallanca et al.}

\begin{document} 

\title{The identification of the optical companion to the binary millisecond pulsar J0610-2100 in the
Galactic field}

\author{
C. Pallanca\altaffilmark{1},
R. P. Mignani\altaffilmark{2,3},
E. Dalessandro\altaffilmark{1},
F.R. Ferraro\altaffilmark{1},
B. Lanzoni\altaffilmark{1},
A. Possenti\altaffilmark{4},
M. Burgay\altaffilmark{4},
E. Sabbi\altaffilmark{5}.
}
\affil{\altaffilmark{1} Dipartimento di Astronomia, Universit\`a degli Studi
di Bologna, via Ranzani 1, I--40127 Bologna, Italy}

\affil{\altaffilmark{2} Mullard Space Science Laboratory, 
University College London, Holmbury St. Mary, Dorking, Surrey, RH5 6NT, UK}
   
\affil{\altaffilmark{3} Kepler Institute 
of Astronomy, University of Zielona G\'ora, Lubuska 2, 65-265, Zielona G\'ora, Poland}

\affil{\altaffilmark{4} INAF-Osservatorio Astronomico di Cagliari,
  localit\`a Poggio dei Pini, strada 54, I-09012 Capoterra, Italy}

\affil{\altaffilmark{5} Space Telescope Science Institute, 3700 San Martin Drive, Baltimore, MD 21218, USA}

\date{June 27, 2012}

\begin{abstract}
We have used deep $V$ and $R$ images acquired at the ESO Very Large Telescope to identify the optical companion
to the binary pulsar \psr,  one of the black-widow millisecond pulsars recently detected by the 
Fermi Gamma-ray Telescope in the Galactic  plane. 
We found a  faint star ($V\sim26.7$) nearly coincident ($\delta r \sim0\arcsec.28$) 
with the  pulsar nominal position. This star is visible only in half of the available images, while it disappears in  the deepest 
ones (those acquired under the best seeing conditions), thus indicating that it is  variable. 
Although our observations do not sample the entire orbital period  ($P=0.28$ d) of the pulsar, we found that the optical
modulation  of the variable star nicely correlates with  the pulsar orbital period and describes a well 
defined peak  ($R\sim25.6$) at $\Phi=0.75$, suggesting a modulation due to the pulsar heating. 
We tentatively conclude that the companion to \psr\ is a heavily ablated very low mass star 
($\approx 0.02\Msun$) that completely filled its Roche Lobe.
\end{abstract} 


\section{INTRODUCTION}
It is generally accepted that millisecond pulsars (MSPs) are formed in binary systems containing a neutron 
star  that is eventually spun up through mass accretion from an evolving companion.
Among these systems those characterized by relatively small eccentricity and very small
mass function (typically the companion mass is only $\mcom \lsim 0.1\Msun$) are classified as ``black-widow'' pulsars
(BWPs). In several cases the pulsar shows eclipses in the radio signal suggesting that the companion is a
non-degenerate, possibly bloated star. In some cases the eclipse of the radio signal is so extended that it implies
a size of the companion larger than its Roche lobe,   suggesting that the obscuring material is plasma released by the
companion because of the energy injected by the pulsar. However since the size of the eclipse depends on the inclination
angle (King et al. 2005), not all BWPs are expected to show eclipses.
As suggested by King et al. (2003), the formation of BWPs needs two phases: a first one in which the 
companion spins-up the neutron star to millisecond periods and a second where the companion is ablated 
by the pulsar. 
While it is difficult to describe the two phases using the same star as a companion to the MSP, 
in globular clusters (GCs), where encounters and exchange interactions are frequent, the white dwarf (WD) companion 
responsible for the pulsar spinning-up can be replaced by a main sequence star  via an exchange interaction.
The following evolution of this newly assembled binary system can cause the 
progressive vaporization of the companion because of the energy injected by the MSP. 
  Since dynamical interactions are less probable in low density environments, BWPs were thought to be mainly generated in
GCs and then ejected in the field. However, the increasing number of BWPs discovered in the Galactic field suggests that 
they must form in the disk as well. 
In this paper we focus on a BWP in the galactic plane: \psr.
 
\psr\ is a MSP with period $P=3.8$ ms and a radio flux at 1.4 GHz of 
$S_{\rm 1.4}= 0.4 \pm 0.2$ mJy discovered during the Parkes High-Latitude pulsar survey 
(Burgay et al.\ 2006, hereafter B06). The period derivative $\dot{P}=1.235\times10^{-20}$s s$^{-1}$ 
implies a characteristic age $\tau=5$ Gyr, a magnetic field $B=2.18 \times 10^8$ G, and a rotational 
energy loss rate  $\dot{E}= 2.3 \times 10^{33}$ 
erg s$^{-1}$, similar to the values measured for other  MSPs.  
\psr\ is in a binary system, with an orbital period of $\sim 0.28$ d. In particular, it is one of 53 
binary MSPs with $P < 10$ ms currently known in the Galactic disk (http://www.atnf.csiro.au/research/pulsar/psrcat/expert.html; 
Manchester et al. 2005).
It is located at a  distance of  $3.5\pm 1.5$ kpc, estimated from its dipersion measure (DM=60.666 pc cm$^{-3}$) 
and the Galactic electron density model of  Cordes \& Lazio (2002). The pulsar has a proper motion of 
$\mu_{\alpha} cos\delta = 7 \pm 3$ mas yr$^{-1}$ and $\mu_{\delta}=11\pm 3$ mas yr$^{-1}$ (B06), which 
implies a transverse velocity of  $228\pm53$ km s$^{-1}$, one of the highest measured for Galactic MSPs. 
The system mass function ($f=5\times10^{-6}$) implies a lower limit of $0.02 \Msun$ for the mass of 
the companion, assuming $1.35 \Msun$ for the pulsar (B06).
Thus, in agreement with the definition above, \psr\ is probably a BWP seen at a low inclination angle 
(in fact no eclipse is detected).

Until a few years ago just two other BWPs were known in the Galactic field, namely PSR B1957+20 
(Fruchter et al. 1988a) and PSR J2051-0827 (Stappers et al.\ 1996a).
But very recently, thanks both to  dedicated surveys of $\gamma$-ray sources and to new blind searches, 
seven new BWPs have been discovered, most of  them having been detected in $\gamma$-rays 
(see Roberts 2011 and references therein).
Also \psr\ has been detected in $\gamma$-rays by the Large Area Telescope (LAT) on board the 
{\em Fermi} Gamma-ray Space Telescope. Based upon positional coincidence with the LAT error box, it was
 initially associated with the $\gamma$-ray source 1FGL\, J0610.7$-$2059 (Abdo et al.\ 2009; 2010). 
The detection of $\gamma$-ray pulsations at the radio period of \psr\ has been recently 
reported and  used to confirm it as the $\gamma$-ray counterpart to the MSP 
(Espinoza et al., in preparation). 
In the X-rays \psr\ has not been detected, neither by the {\em ROSAT} All Sky Survey 
(Voges et al.\ 1999), nor by {\em Swift} (Marelli, private communication).

Studying the optical emission properties of binary MSP companions is important to better constrain the orbital
parameters and to clarify the evolutionary status of these systems and then to track back 
their history and characteristic timescales. 
In spite of their importance, only two optical companion to BWPs in the Galactic field have been 
detected to date (Fruchter et al. 1988b,  Reynolds et al. 2007, and Van Kerkwijk et al.2011; 
Stappers et al.\ 1996b,  Stappers et al.
1999 and Stappers et al. 2000). 
The binary MSP PSR B1957+20 is the first discovered BWP and one of the best studied members
of this class. The optical companion to PSR J1957+20 was identified by Kulkarni
et al. (1988), while subsequent observations found the companion to vary by a
fraction of 30\%-40\% in flux over the course of the orbital period (Callanan et
al. 1995). Reynolds et al. (2007), modelling the light curve over all the orbital period,
constrained the system inclination $63 ^{\circ}<i<67^{\circ}$ and the filling factor of the
Roche Lobe ($0.81<f<0.87$). Moreover, they ruled out the possibility that the companion is a white dwarf, 
suggesting that most probably is a brown dwarf.
 A recent spectroscopic analysis, combined with the knowledge of the inclination
angle inferred from models of the light curve, suggested that the PSR B1957+20 is
massive with $M_{\rm PSR}=2.4\Msun$ ($M_{\rm PSR}>1.66\Msun$ being conservative; van Kerkwijk
et al. 2011).
The optical companion to binary PSR J2051-0827 was identified by Stappers et al. (1996b).
They found that the amplitude of the companion's light curve was at least 1.2 mag, and that the
variation was consistent with the companion's rotating synchronously about the pulsar and one
side being heated by the impinging pulsar flux.
In  subsequent works it has been possible to study the entire lightcurve, measuring amplitudes of
3.3 and 1.9 magnitudes in the $R$-band and $I$-band respectively. The companion star has 
been modelled by a gravitationally distorted
low-mass secondary star which is irradiated by the impinging pulsar wind. The resulting
best-fit model is of a Roche lobe filling companion star which converts approximately 30\% of
the incident pulsar spin down energy into optical flux (Stappers et al., 2000).
 
Here we present  the  first identification of  the optical companion to \psr, from data acquired at the 
ESO {\em Very Large Telescope} (\vlt). The observations and data analysis are described in Sect. 2, 
while the results are presented  in Sect. 3 and discussed in Sect. 4.

\section{OBSERVATIONS AND  DATA ANALYSIS}\label{Sec:analysis}
The photometric data set used for this work consists of a series of ground-based optical images 
acquired with the FOcal Reducer/low dispersion Spectrograph 2 (FORS2) mounted at the ESO-VLT.
We used the Standard Resolution Collimator, with a pixel scale of 0.25\arcsec /pixel and a field 
of view (FOV) of $6.8'\times6.8'$.  In order to obtain deep images free from the blooming due to 
heavy saturation of bright stars, which can significantly limit the search for faint objects, 
all the brightest stars in the FOV have been  covered with occulting masks.

Six short acquisition images (of 5 s each) and a total of 29 deep images in the $V_{BESSEL}$ and 
$R_{SPECIAL}$ bands ($V$ and $R$ hereafter)  were collected during six nights, from mid 
December 2004 to the beginning of January 2005 (see Table \ref{dataset}), under program 
$074.D-0371(A)$ (PI: E. Sabbi). 
These data allow us to sample  $\sim 25\%$ of the orbital period in $V$ and less than $40\%$ of 
it in $R$ (see column 4 in Table \ref{dataset}).

By following standard reduction procedures, we corrected the raw images for bias and flat-field.
In particular, in order to obtain high-quality  master-bias and  master-flat images, we selected  
a large number of  images obtained during each observing night  and, for each filter, we properly 
combined them by using the tasks  {\tt zerocombine} and {\tt flatcombine} in the IRAF package 
{\sc ccdred}. The calibration files thus obtained have been applied to the raw images by using 
the dedicated task  {\tt ccdproc}.

Based on the  Word Coordinate System (WCS) of the images, we approximately located the  pulsar 
position and decided to limit the photometric analysis to a region of 500 pixel $\times$ 500 pixel 
($\sim$ 125\arcsec $\times$ 125\arcsec) centered on it.

We carried out the photometric analysis by using  {\sc daophot} (Stetson 1987, 1994).
We modeled the point spread function (PSF) in each image by using about forty bright, isolated 
and not saturated stars.
The PSF model and its parameters have been chosen by using the  {\sc daophot} {\tt PSF} routine  on the basis 
of a Chi-square test,  with a Moffat function (Moffat 1969) providing the best fit to the data in 
all cases.

Since our purpose is to detect the faintest stars in the field, we selected the three images 
obtained under the best seeing conditions ($\sim0.6\arcsec$) in both filters and we combined 
them using the IRAF task  {\tt imcombine} in order to obtain a high signal to noise (S/N) 
master frame  to be used as a reference for the object detection. 
This was done by using the  {\sc daophot} {\tt FIND} routine and imposing  a detection limit of $3 \sigma$.

Finally,  by using {\tt allframe} (Stetson 1987, 1994) we forced the object detection and  PSF fitting 
in each single image adopting the star positions in the master frame as reference. This procedure 
also allowed us to achieve an improved determination of the  star centroids and a better 
reconstruction of the star  intensity profiles.

{\it Photometric calibration:} 
For a straightforward comparison with theoretical models able to provide information about 
luminosity and temperature of the companion star, we decided to calibrate the instrumental 
magnitudes to  the standard Johnson photometric system.
To this aim, we first derived the calibration equation for ten standard stars in the field 
PG0231 (Stetson 2000), which has been observed with FORS2 during a photometric night 
(December 17, 2004) in both $V$ and $R$ under  a calibration program.
To analyze the standard star field we used the {\sc daophot} {\tt PHOT} task and performed aperture 
photometry with a radius $r=14$ pixels beyond wich the  contribution of the PSF wings  to the 
star  intensity profile becomes negligible.
We then compared the obtained magnitudes with the standard  Stetson catalog available on the 
 CADC web  site\footnote{http://cadcwww.dao.nrc.ca/community/STETSON/standards/}. 
The resulting calibration equations are $V=v-0.092(v-r)+27.79$ and $R=r-0.019(v-r)+27.97$, where 
$v$ and $r$ are the instrumental magnitudes.
The color coefficient is very small for the $R$ band, while it could be not negligible in the
$V$ band.

{\it Astrometry:}  Since the pulsar timing position is known with a very high precision, obtaining  
accurate astrometry is a critical requirement in searching for 
the optical companions to binary MSPs.
For this reason, particular care has been devoted to obtain a  very good astrometric solution.
Since most of the astrometric standard stars are saturated in our catalog, we used the short $R$-band to this purpose.
As a first step of our procedure, we  registered the  pixel coordinates of this image onto 
the absolute coordinate system through the cross-correlation  of about one hundred  primary 
astrometric standards from Guide Star Catalog II (GSCII - STScI, 2001; Lasker et al. 2008), 
by using CataXcorr\footnote{CataXcorr is a code aimed at cross-corralating catalogues and finding 
astrometric solutions, developed by P. Montegriffo at INAF - Osservatorio Astronomico di Bologna. 
This package has been used in a large number of papers of our group in the past 10 years.}.
We then used  about fifty  not saturated
 stars in common between the reference and our general catalogs as 
secondary astrometric standards.
 At the end of the procedure the typical accuracy of the astrometric solution was $\sim0\arcsec.2$ in both 
right ascension ($\alpha$) and declination ($\delta$).

\section{THE COMPANION TO PSR J0610-2100}
In order to  identify the companion to \psr\ we  first searched for objects with 
coordinates compatible with the nominal PSR position: $\alpha= 06^h 10^m 13^s.59214(10)$, 
$\delta=-21^{\circ} 00' 28''.0158(17)$ at Epoch MJD=53100 (B06). 
Since the epoch of observations  is within less than one year from the epoch  of the reference radio 
position, we neglected the effect of proper motion, which is much smaller than the 
 accuracy of the astrometric solution of the FORS2 images.

\begin{figure*}
\includegraphics[width=150mm]{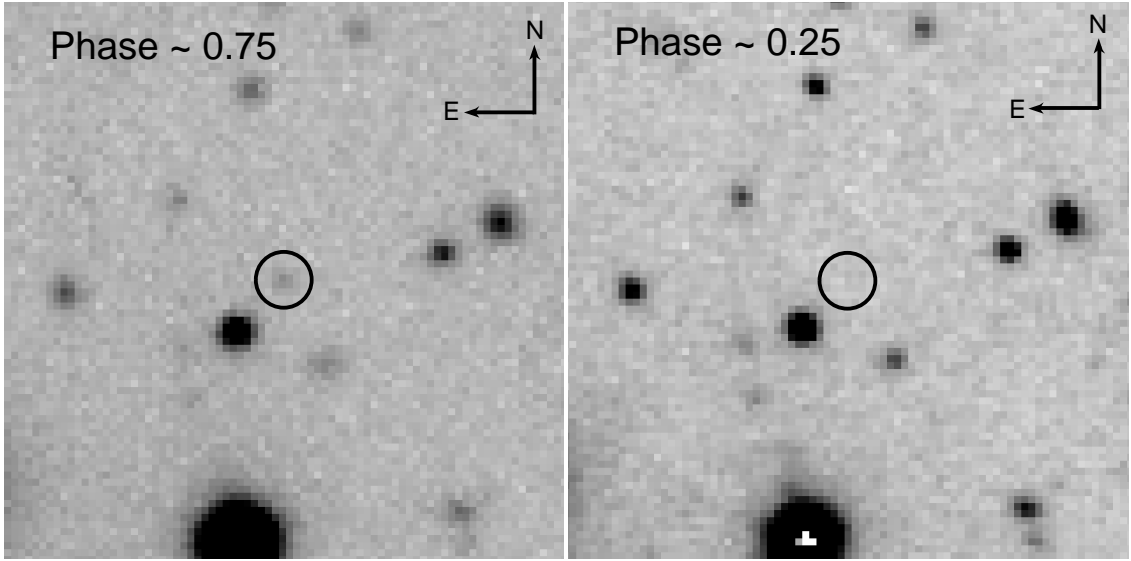}
  \caption{$R$-band images of  the $20\arcsec\times20\arcsec$ region around the nominal position of \psr, at two different epochs
  corresponding to the orbital phases $\Phi=0.75$ (left panel) and $\Phi=0.25$ (right panel). Each image is the average of 3
  images. The circle of radius $1\arcsec$ marks the pulsar position. A star is clearly visible in the left panel,
  while it vanishes in the right panel. Note that the images at $\Phi=0.25$ (where the star is undetected) have been obtained 
  under the best seeing conditions   ($FWHM=0.6\arcsec$).}
\label{map}
\end{figure*}

\begin{figure*}
\includegraphics[width=150mm]{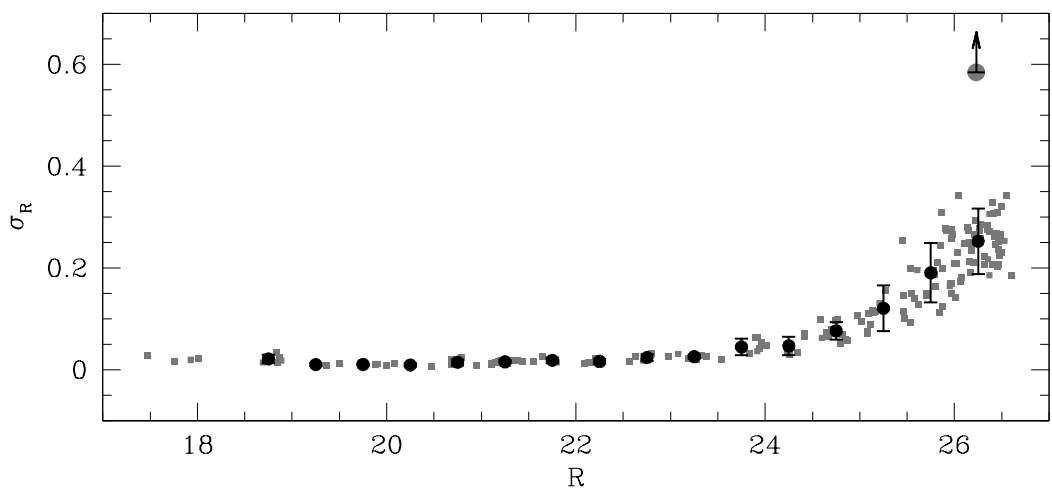}
  \caption{Frame to frame scatter (gray points) for the 173 stars identified in a $125 \arcsec \times125 \arcsec$ region 
  around the nominal PSR position,  as a function of the $R$ magnitude.
   The standard deviation $\sigma _R$ has been computed by using all the available images. 
   Black circles and the corresponding error bars are the mean and the standard deviation values 
    in 0.5 magnitude bins, respectively. 
      The mean magnitude ($R\sim26.2$) and standard deviation ($\sigma_R\sim0.6$) of the companion star to PSR J0610-2100
     (large grey circle) likely represent lower limits to the true values, because they have been computed adopting $R=27$
     (the $R$-band detection limit) in all images where the star was not visible.
 }
\label{sigma}
\end{figure*}

A first visual inspection of the pulsar region clearly shows that only one star lies within 
a couple of arcseconds from the MSP radio position: it is located at $\alpha= 06^h 10^m 13^s.58$ 
$\delta=-21^{\circ} 00' 27''.83$, just $0.\arcsec28$ from \psr. 
Thus, from positional coincidence alone, we found a very good candidate  companion to \psr. 
 Note that the chance coincidence probability \footnote{The chance coincidence probability 
is calculated as $P=1-\exp(-\pi \sigma R^2)$ where $\sigma$ is the
stellar density of stars with similar magnitude to the candidate companion and $R$ is the accuracy of the astrometric solution.} 
that a star is located at the pulsar position is only $P=0.0007$. Hence, this star 
is the companion to \psr\ with a probability of $\sim 99\%$.
Interestingly enough, this star was not present in the  master object list obtained 
from the stacking of the $R$-band images
because it is visible in only half of the images, while it completely 
disappears in the others (see Figure \ref{map}). 
We performed a detailed photometric analysis of this star for measuring its magnitude in as many 
images as possible, and we found that it is not  detected in the  deepest $R$ images obtained 
under the best seeing conditions ($FWHM\sim0.6\arcsec$).
In summary, we were able to measure the magnitude of the star in only 12 images (10 in the 
$R$-band and 2 in the $V$-band), finding significant variations: $\delta R \sim 1$ mag, from 
$R=25.3 \pm 0.1$ to $R=26.3 \pm 0.2$, and $\delta V \sim 0.5$ mag, from $V=26.7 \pm 0.2$ to $V=27.2 \pm 0.2$. 
In the remaining images the star magnitude is below the  detection threshold
($R=27 \pm 0.3$ and $V=27.3 \pm 0.3$),  thus suggesting a much more pronounced optical variation.
Considering the entire data set,  the object's photometry  shows a quite large scatter, 
significantly ($>5\sigma$) larger than that computed for stars of similar magnitude in the same 
FOV (see Figure \ref{sigma}). 
These findings confirm that this is a variable object near the detection limit of our sample.

\begin{figure*}
\includegraphics[width=150mm]{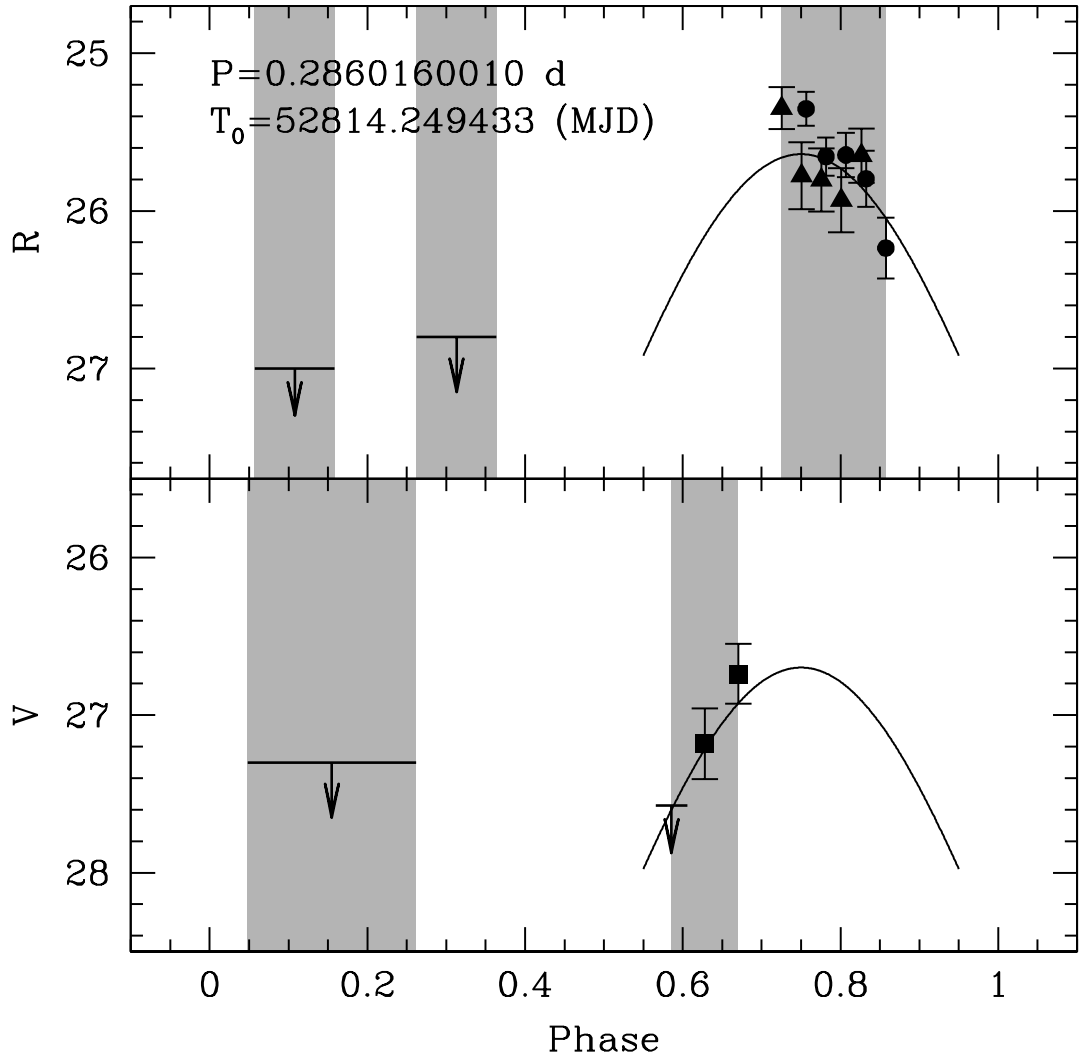}
  \caption{The observed light curve of the companion to \psr\ folded with the orbital period (P) of the pulsar using the  
  reference epoch ($T_0$) known from radio observations (B06).
  The different symbols represent images obtained in different nights. The horizontal lines with arrows are the estimated 
  magnitude upper-limits for the images where
  the star is  below the detection threshold. The gray regions correspond to the fraction of the orbital phase sampled by 
  each observations.   The solid black line is a first-guess modeling of the light curve by means of a simple mode.}
\label{lc}
\end{figure*}

In order to establish a firm connection between this star and the pulsar we computed the $V$ 
and $R$ light curves folding each  measurement with the orbital period ($P=0.2860160010$ d ) 
and the  ascending node ($T_0=52814.249433$)  from the radio ephemeris (B06).
As shown in Figure \ref{lc}, although the available data do not allow a complete coverage of the 
orbital period, the  flux modulation of the star  nicely correlates with the pulsar orbital 
phase. The available data are consistent with  the rising (in the $V$ band) and the decreasing 
(in $R$-band) branches of a light curve with a peak at $\Phi=0.75$.
This is the typical behaviour expected when the surface of the companion is heated by the pulsar 
flux  and the orbital plane  has a sufficiently high inclination angle. 
In fact, in this configuration a light curve with a maximum at $\Phi=0.75$ (corresponding to the 
 pulsar inferior conjunction, when the companion surface faces the observer) and a minimum at 
$\Phi=0.25$ (corresponding to the  pulsar superior conjunction) is expected.
Indeed  the star is not detectable at the  epochs corresponding to the orbital phases where the 
luminosity minimum is predicted. 

Based on all these pieces of evidence we propose the identified variable star as the companion 
to the pulsar; according to previous papers (see Ferraro et al. 2001, 2003; Cocozza. et al 2006 
and Pallanca et al. 2010), we name it \com.

\begin{figure*}
\includegraphics[width=150mm]{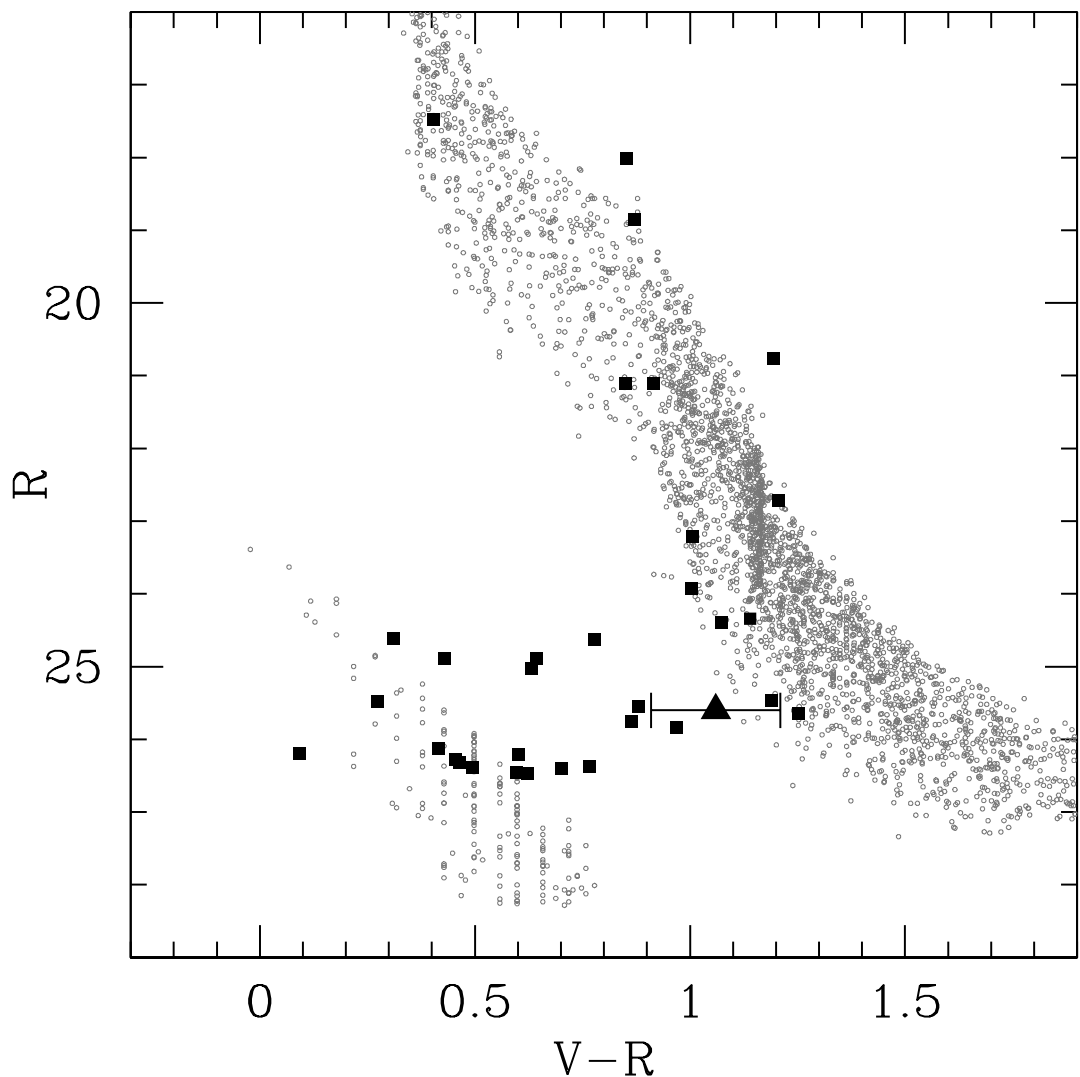}
  \caption{($R$, $V-R$) color-magnitude diagram for \com\ (large black triangle) and for the objects detected in a 
  circle of about $30\arcsec$ around its nominal position (black squares).  
  \com\ is located at $V=26.7$ and $R=25.6$, corresponding to the values  estimated from the first-guess light-curve at its
  maximum ($\Phi=0.75$). The error bar corresponds to the photometric  error at those magnitude values.
  The open gray circles represent the Galaxy disk population obtained with the
  Besancon model (Robin et al. 2003) in the direction of the pulsar and for a distance between 2 and 6 kpc.}
\label{cmd}
\end{figure*}

Since the available $V$ and $R$ measurements are mainly clustered toward the maximum of the 
emission, but do not allow to precisely determine it, we used  a simple  sinusoidal 
function\footnote{ Although this assumption is not supported by a physical reason, it provides a first estimate of
the magnitude and colour of COM 
J0610-2100 at maximum.} to obtain a first-guess modeling of the light curve.
In the following analysis we
will use these values instead of the mean magnitudes over the entire orbital period. In fact,
while the latter would be more appropriate in general, they are not available in this case
 because of the incomplete sampling of the period. Also note that during the calibration 
procedure the color term entering the equations has been computed as the difference between the 
average value of the available $V$ and $R$ instrumental magnitudes. While this is strictly correct 
for non variable stars, in the case of \com\ it could have introduced an error in the 
estimated magnitudes.  Since in the calibrating equations the coefficients of the color terms are very small, 
especially in the $R$-band, this uncertainty should be negligible.
The resulting magnitudes of \com\ at maximum are $R=25.6$ and $V=26.7$.
Figure \ref{cmd} shows the position in the ($R$, $V-R$) color-magnitude diagram (CMD)  of 
\com\     and the stars detected within  $30\arcsec$ from it.
For the sake of clarity a simulation of the Galactic disk population in the direction of 
\psr\ computed with the Besancon Galaxy model (Robin et al. 2003) is also shown.
As can be seen, \com\ is located at a slightly bluer  color with respect to the 
reference main sequence, thus suggesting that it probably is a non degenerate, low mass, 
swollen star. Indeed similar objects have been previously identified in Galactic globular 
clusters (see Ferraro et al. 2001, Edmonds et al 2002, Cocozza et al. 2006 and 
Pallanca et al. 2010).

\section{DISCUSSION}
We have determined the physical parameters of \com\ from the comparison of 
its position in the CMD (Figure \ref{cmd}) with a reference zero age main sequence,
assuming an interstellar extinction of $E(B-V)\sim0.074$\footnote{ from NED, 
Nasa/ipac Extragalactic Database - Galactic Extinction Calculator available at the 
web site {\it http://ned.ipac.caltech.edu/forms/calculator.html}.} and the typical 
metallicity of the Galactic disk ($Z=0.02$).
The resulting effective temperature and bolometric luminosity of the star are 
$T_{eff}\sim3500$ K and $L_{bol}\sim 0.0017 L_{\odot}$ respectively, with a conservative 
uncertainty of $\pm 500$ K and $\pm0.0001L_{\odot}$.
Under the assumption that  the optical emission of \com\ is well reproduced by a blackbody (BB), it is 
possible to derive its radius: $R_{\rm BB}\sim 0.14R_{\odot}$.
However, since the companion to a BWP is expected to be affected by the tidal distortion 
exerted by the pulsar and to have filled its Roche Lobe, the dimension of the Roche Lobe 
might be a more appropriate value  (i.e., see PSR J2051-0827; Stappers et al. 1996b and Stappers et al. 2000).
According to Eggleton (1983) we assume: 
\begin{displaymath}
\rrl \simeq \frac{0.49q^{\frac{2}{3}}}{0.6q^{\frac{2}{3}}+\ln \left(1+q^{\frac{1}{3}}\right)}  
\end{displaymath}
where $q$ is the ratio between the companion and the pulsar masses ($\mcom$ and 
$\mpsr$, respectively).
This relation can be combined with the PSR mass function 
$f(i,\mpsr,\mcom)=(\mcom\sin i)^3/(\mcom+\mpsr)^2$ 
by assuming a  NS mass $\mpsr=1.5 \Msun$ (as recently estimated for recycled pulsars by Ozel et al. 2012; 
see also Zhang et al. 2011 and Kiziltan et al. 2011),
thus yielding  $\rrl(i)\sim0.24-0.47 R_{\odot}$, depending on the inclination angle 
($i$) of the orbital system. These values are about  1.7-3.4 times larger than $R_{\rm BB}$. 
In the following discussion we assume  the value of the Roche Lobe as a measure of the 
size of \com\ and we discuss how the scenario would change by using $R_{\rm BB}$ 
instead of $\rrl$.
While these assumptions trace two extreme possibilities, the situation is probably in the midway. 
In fact, in the case of a completely filled Roche Lobe, the mass lost from the companion should 
produce some detectable signal in the radio band (unless for very small orbital inclinations) 
and ellipsoidal variations could be revealed in the light curve (unless the heating from the pulsar is dominating).

Under the assumption that the optical variation shown in Figure \ref{lc} is mainly 
due to irradiation from the MSP reprocessed by the surface of \com\, we can 
estimate how the re-processing efficiency depends on the inclination angle and, hence, 
on the companion mass.
To this end, we compare the observed flux variation ($\Delta F_{obs}$)  between the 
maximum ($\Phi=0.75$) and the minimum ($\Phi=0.25$) of the light curve, with the 
expected flux variation ($\Delta F_{exp}$) computed from the rotational energy loss  
rate ($\dot{E}$).
Actually, since we do not observe the entire light curve, $\Delta F_{obs}$ can just 
put a lower limit to the reprocessing efficiency.
Moreover, since these quantities depend on the inclination angle of the system 
(see below) we can just estimate the reprocessing efficiency as function of $i$.

At first we have to convert the observed magnitude variation into a flux. We limited 
our analysis to the $R$ band since we have more observations and a more reliable 
sampling of the light curve. 
At maximum ($\Phi=0.75$) we assume $R=25.6$, and between $\Phi=0.75$ and $\Phi=0.25$ 
we estimate an amplitude variation $\Delta R \gapp 1.5$.
Hence we obtaine $\Delta F_{obs} \sim 1.88 \times 10 ^{-30}$ erg s$^{-1}$ cm$^{-2}$ Hz$^{-1}$ 
and considering the filter width 
($\Delta \lambda = 165$ nm) we have $\Delta F_{obs} \sim 3.4 \times 10 ^{-15}$ 
erg s$^{-1}$ cm$^{-2}$.

\begin{figure*}
\includegraphics[width=150mm]{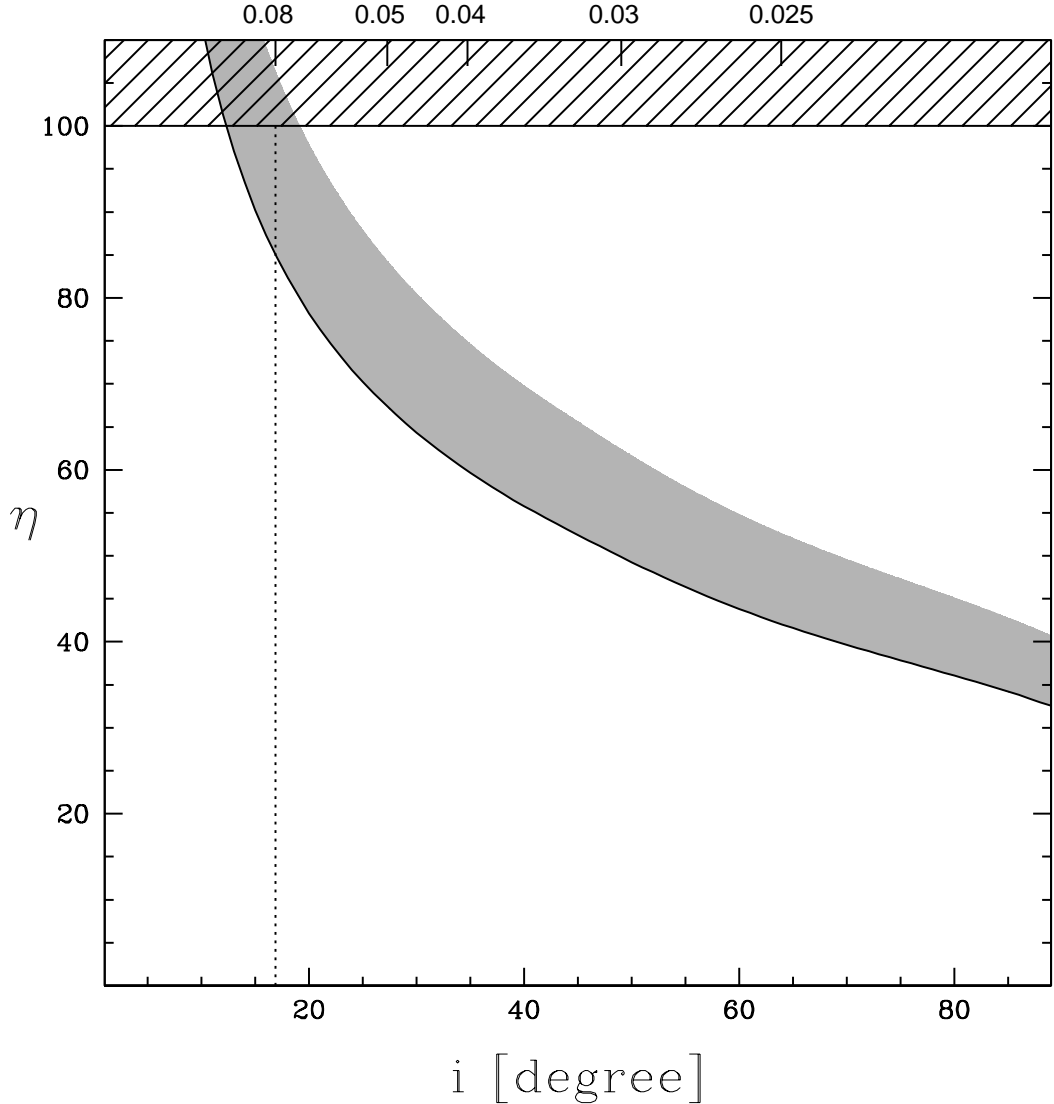}
  \caption{The reprocessing efficiency for isotropic emission ($\eta$) as a function of the inclination angle ($i$) calculated assuming
   $\mpsr=1.5 \Msun$ and $\rcom=\rrl(i)$. The corresponding values for the companion mass in units of $\Msun$ are reported on
   the top axis.
   The solid line corresponds to the lower limit of the amplitude of variation  ($\Delta R =1.5$), while the gray region to
   values of $\Delta R$ up to 3, that should be appropriate if considering the entire light curve (i.e, see PSR J2051-0827; Stappers et al. 1996b). 
   The shaded area marks the region of the diagram where only anisotropic emission of the re-processed flux is admitted (in case of
   isotropy, in fact, the efficiency would be unphysical: $\eta>100\%$).  
  The dotted line marks  the physical limit for core hydrogen burning  stars (i.e., objects with masses $\ge 0.08\Msun$).}
\label{repr}
\end{figure*}

On the other hand, the expected flux variation between $\Phi=0.75$ and $\Phi=0.25$ 
is given by 
\begin{displaymath}
\Delta F_{exp}(i)= \eta \frac{\dot{E} }{A^2}  \rcom^2  \frac{1}{4\pi d^2_{\rm PSR}} \varepsilon (i)
\end{displaymath}
where $\eta$ is the re-processing efficiency under the assumption of isotropic 
emission, $A$ is semi-major axis of orbit which depends on 
the inclination angle, $\rcom$ is the radius of the companion star,  which we 
assumed  to be equal to $\rrl(i)$, $\dpsr$ is the distance of pulsar  (3.5 kpc \footnote{In these calculations 
we adopted a distance of 3.5 kpc, while we discuss below how the scenario changes  
by varying the distance between the range of values within the quoted uncertainty.}) and
$\varepsilon (i)$ is the fraction of the re-emitting surface visible to the 
observer\footnote{ In the following we assume $\varepsilon (i)=i/180$. In fact, for a face-on configuration 
($i=0^{\circ}$) no flux variations are expected, while for an edge-on system ($i=90^{\circ}$) 
the fraction of the heated surface that is visible to the observer varies between
0.5 (for $\Phi=0.75$) to zero (for $\Phi=0.25$).}.
By assuming $\Delta F_{obs}=\Delta F_{exp}(i)$ between $\Phi=0.75$ and $\Phi=0.25$,
we can derive a relation linking the re-processing efficiency and the inclination 
angle. The result is shown in Figure \ref{repr}.
The absence of eclipses in the radio signal allows us to exclude very high 
inclination angles.
As shown in Figure \ref{repr}, pulsar companions  with stellar mass above the physical 
limit for core hydrogen burning star 
(i.e., with $M\ge0.08\Msun$) necessarily imply a non-isotropic emission mechanism 
of the pulsar flux 
(otherwise a larger than $100\%$ physical efficiency would be required). 
On the contrary a re-processing efficiency between 40\% and 100\% is sufficient for 
less massive companions and intermediate inclination angles.
 
Taking into account the uncertainty on the pulsar distance, only re-processing 
efficiencies larger than $\sim60\%$ for inclination angles in excess to $~50^{\circ}$ and 
companion masses lower than $\sim0.03\Msun$ are allowed in the case of the distance
upper limit (5 kpc). Instead, in case of a closer distance $\eta$ decreases for all 
inclination angles, thus making acceptable also companion stars with masses larger 
than $0.08 \Msun$. For instance, for companions masses between $0.08$ and $0.2 \Msun$ and 
a 2 kpc distant pulsar, the re-processing efficiency ranges between $30\%$ and $60\%$ 
for any value of $i$.

 The observed optical modulation can be reproduced considering a 
system seen at an inclination angle of about $60^{\circ}$,  with a very low mass 
companion ($\mcom\sim0.02\Msun$) that has filled its Roche Lobe, and a 
re-processing efficiency of about $50\%$. 
On the other hand, if we use $R_{\rm BB}$ instead of the $\rrl$, the efficiency 
becomes larger than 100\% for every inclination angle, and the only possible
scenario would be that of an anisotropic pulsar emission. However, even with this 
assumption it is very difficult to obtain an acceptable value for $\eta$. 
This seems to confirm that $R_{ \rm BB}$ is too small to provide a good estimate of 
the star physical size. 
Forthcoming studies will allow us to better constrain the system parameters.

\section{ACKNOWLEDGEMENT}
We thank the Referee Scott Ransom for the careful reading
of the manuscript and the useful comments. This research is part of the project COSMIC-LAB funded by the European Research Council 
(under contract ERC-2010-AdG-267675). We thank G. Cocozza for useful discussions.

\newpage
\begin{table}
\begin{center}
\begin{tabular}{|l|l|l|l|c|}
\hline
FILTER &  N $\times$ $t_{exp}$ & Night & PSR-Phase & Detection \\
\hline
\hline
$V_{BESSEL}$ & 6 $\times$ 1010 s & 17/12/04 & 0.04-0.26 & NO\\ 
$V_{BESSEL}$ & 3 $\times$ 1010 s & 20/12/04 & 0.58-0.67 & YES$^1$\\ 
\hline
$R_{SPECIAL}$ & 5 $\times$ 590 s & 14/12/04 & 0.73-0.83 & YES\\ 
$R_{SPECIAL}$ & 5 $\times$ 590 s & 21/12/04 & 0.06-0.16 & NO\\ 
$R_{SPECIAL}$ & 5 $\times$ 590 s & 05/01/05 & 0.26-0.36 & NO\\ 
$R_{SPECIAL}$ & 5 $\times$ 590 s & 06/01/05 & 0.76-0.86 & YES\\ 
\hline
\end{tabular}
\end{center}
\caption{$^1$ Detected in only 2 images out of 3}
\label{dataset}
\end{table}

\newpage

\newpage

\newpage

\newpage

\newpage

\end{document}